# Effect of particle oxidation, size and material on deformation, bonding and deposition during cold spray: a peridynamic investigation

Baihua Ren, Jun Song[1]

*Department of Mining and Materials Engineering, McGill University, Montréal, QC H3A 0C5, Canada*

## Abstract

Cold spray (CS) has emerged as an important additive manufacturing technology over the past decade. This study investigates the effect of oxide layers on the CS process, focusing on the deformation behavior of copper (Cu) and iron (Fe) particles upon collision with a matching substrate. Using a peridynamics-based approach, we examine the effects of oxide thickness, particle size, and particle/substrate material on material deformation and oxide fracture processes. Our results show that thicker oxide films restrict particle deformation, delay oxide discontinuities and material jetting, and increase the critical velocity required for metal–metal contact. Larger particles, despite uniform deformation across sizes, require lower velocities to initiate jetting and oxide separation because of their higher kinetic energy, leading to metallurgical bonding at lower velocities. Soft-to-soft impacts induce oxide film cracking at lower velocities, resulting in larger interface areas and more oxide-free contact zones, thereby reducing the critical velocity. Furthermore, the volume of residual oxide has a power-law relationship with the particle size, indicating that the oxide-cleaning ability of the particles affects the critical velocity. This study highlights the importance of oxide deformation and fracture during CS processes and provides valuable insights into the breakage and removal of oxides and subsequent metallic bond formation. These findings offer beneficial new knowledge for the rational design and optimization of CS processes.

[1] Author to whom correspondence should be addressed. E-Mail : jun.song2@mcgill.ca, Tel : (+1) 514-398-4592

## 1. Introduction

Cold spray (CS) has attracted significant attention as a major advancement in material deposition and as a promising additive manufacturing (AM) process [1-4]. This technology helps mitigate the generation of high-temperature-induced residual stresses that are common in conventional AM techniques [5]. In CS [1,3,6-8], micron-sized particles are accelerated in a high-pressure gas stream through a DeLaval-type nozzle [2,9,10] before reaching a substrate. These supersonic feedstock powders impact and adhere to the substrate through intense plastic deformation. Unlike conventional thermal spray methods, which cause severe oxidation, as well as microstructural and phase transformations due to thermal effects [11,12], CS causes minimal heating of the particles prior to impact, with impact velocities ranging from 200–1200 m/s. The bonding in CS originates from particle deformation during impact, a phenomenon that is complexly affected by the specific impact conditions and various powder properties, as detailed in a previous study [13].

The optimization of the coating microstructure and properties, tailored for each combination of substrate and coating materials, necessitates a series of spray experiments and corresponding characterizations. However, the absence of a comprehensive understanding of the mechanisms governing bonding and coating formation can render these optimization processes excessively time-consuming, costly, and, in some cases, unfeasible. The endeavour to fine-tune CS for specific materials has driven investigations into bonding mechanisms, drawing parallels between key characteristics of bonding in CS and processes such as explosive welding or shock wave powder compaction [14]. According to prevailing bonding theories [15], all metals and alloys can form bonds when their clean surfaces come into contact within the range of interatomic forces. Nevertheless, the majority of metal surfaces undergo oxidation in ambient air, resulting in the development of a thin film that may gradually thicken over time [16]. This native layer is recognized as a barrier to metallurgical bonding and necessitates breaking and displacement during cold spraying to establish fresh metal–metal interfacial contacts. During cold spraying, plastic deformation occurring at high strain rates can lead to partial removal of the native oxide layer. The substantial kinetic energy of the particles before they impact the substrate induces severe local plastic deformation in both the particle and the substrate, resulting in the formation of material jets at the particle–substrate interface [13,14]. However, metallic oxides, which are inherently brittle, are incapable of undergoing substantial plastic deformation. It is believed that the significant plastic deformation involved in jetting, which disrupts the surface films into numerous debris, is responsible for creating gaps that expose fresh material [17-21]. Therefore, achieving the appropriate particle impact velocity in CS is crucial, as it must be of sufficient magnitude to disrupt the thin native oxide layer present at the solid interfaces and facilitate contact between the two clean surfaces [17,18,20,22-27]. If the deformation level falls short, remnants of debris may persist at the interface, impeding local intimate contact [28,29]. The minimum particle velocity at which bonding occurs is termed the critical velocity and is an important parameter in cold spraying [13,14,30]. This velocity is dependent not only on the properties of the sprayed material and substrate but also on the powder size, morphology, composition, and oxidation conditions of both the particle and the substrate [31,32].

Among all the variables mentioned, a major factor believed to significantly influence the critical velocity and final coating quality in CS is the native oxide layer [33]. Experiments have shown that for severely oxidized powders, the critical velocity is often determined by the oxides on the powder surface rather than by the material properties [34]. However, direct experimental observation of the oxide layer removal process during CS is challenging [35]. Computational methods, on the other hand, offer a cost-effective means to assess the effect of the native oxide layer on deposit quality, and consequently, many numerical studies have been performed to investigate the role of oxides in the CS process [11,20,26,29,33,36,37]. In these numerical efforts, the finite element method (FEM) has been the most extensively utilized method for studying particle deformation and coating characteristics in CS [24,38-40]. However, the dynamic nature of the CS process often leads to instabilities and convergence issues in FEM simulations due to high strain rates and extreme plastic deformation [11,33]. Moreover, traditional FEM models face challenges in treating damage, as their formulations are based on continuum mechanics, which are not suitable for discontinuities such as cracks. Consequently, only a limited number of studies utilizing FEM approaches have incorporated the effects of oxide layers [20,26,29]. To address the failure behavior of the oxide layer, specialized techniques such as the element deletion method must be utilized, potentially leading to a violation of the conservation of mass and energy. Furthermore, in many FEM simulations, the thickness of the considered oxide layers is on the order of several hundreds of nanometers [11,20,26,29], which is tens of orders of magnitude greater than the actual thickness of the native oxide film, owing to computational limitations. Meshless methods such as smoothed particle hydrodynamics (SPH) have also been utilized [33,41,42], demonstrating advantages in simulating particle impact problems by mitigating significant mesh distortion encountered in mesh-based methods. However, SPH also faces singularity problems due to discontinuities resulting from oxide damage [43]. Only a limited number of SPH studies have directly modeled oxide layers, with a restricted thickness of hundreds of nanometers [33]. Moreover, on the other side of the length scale spectrum, molecular dynamics (MD) [44-46] simulations have been employed to study high-velocity impacts [36,47-51]. However, the complexity of implementing oxide layers in MD simulations limits their application. Only a few studies have investigated the effects of a brittle layer on metallurgical bonding [36,37]. While MD simulations provide valuable physical insights into the adhesion process, their feasibility is hampered, as the simulated particles must be orders of magnitude smaller than the actual particles in the CS. This limitation undermines the utility of MD as a predictive tool. These challenges have prompted researchers to adopt alternative numerical simulation approaches for studying CS parameters.

In a previous study by the authors, a peridynamics (PD)-based approach was employed to simulate copper (Cu) particles impacting a Cu substrate [52]. The PD results successfully reproduced crucial factors such as splat deformation, the coefficient of restitution, and the onset of jetting in both the substrate and the particle at varying impact velocities, which is consistent with the experimental findings. These results demonstrated that PD simulations can realistically portray actual CS conditions and provide accurate descriptions of the deformation and damage processes involved [52]. PD offers several advantages over FEM, especially when dealing with the complexity of oxide layers in CS processes. Unlike FEM, PD can naturally accommodate discontinuities such as cracks and fractures without the need for additional criteria or methods [53,54]. This inherent capability makes PD well suited for simulating the failure and removal of oxide layers during CS processes. The meshless implementation of PD also alleviates the instability and convergence issues that plague FEM simulations at high strain rates and extreme plastic deformations [5]. On the basis of the PD approach established in our previous work [52], here, we investigated the deformation behaviors of Cu and iron (Fe) particles upon impact on a matched substrate in the presence of oxide.

In the present study, first, the evolution and deformation behaviors of the oxide film on a Cu particle were comprehensively investigated, as the oxide thickness varies from 2.5 nm to 60 nm. Detailed analyses were performed on the effects of the oxide film thickness on the interfacial stress, temperature, material jetting, and particle deposition. The investigation was then extended to include iron (Fe) particles to compare the ability of softer (Cu) and harder (Fe) particle materials to induce oxide fragmentation and removal, thus elucidating the effect of the particle material. Finally, a qualitative prediction is developed to relate oxide removal to particle size in the context of assessing the critical velocity for both Cu and Fe. Finally, the main findings are summarized, and the implications of our results for CS are discussed.

## 2. Methodology

In this section, the material model, PD simulations and oxide damage model are described in detail below. All the PD analyses in this research are conducted via the open-source code Peridigm [55].

### 2.1. Material model

This paper concentrates on Cu and Fe particles impacting matching substrates. We utilize the Johnson–Cook (JC) plasticity model as the constitutive plasticity model, which incorporates the influences of strain, strain rate, and temperature. The Von-Mises equivalent stress derived from the JC model is expressed as follows [56]:

$$\sigma = [A + B\varepsilon^n][1 + C\ln\dot{\varepsilon}^*][1 - (T^*)^m] \tag{1}$$

$$T^* = \frac{T - T_{ref}}{T_m - T_{ref}} \tag{2}$$

Here, $\sigma$ represents the flow stress, $\varepsilon$ represents the equivalent plastic strain (PEEQ), $\dot{\varepsilon}*$ represents the equivalent plastic strain rate normalized by the reference strain rate, $T_{ref}$ represents the threshold temperature allowing thermal softening of the particles, and $T_m$ represents the melting temperature of the metal. Constants *A*, *B C*, *n* and *m* are experimentally derived, and *T* denotes the initial temperature of the particle, which is typically set to room temperature. Table 1 provides an overview of the material properties. These material properties are crucial for understanding the behavior of Cu and Fe particles during impact simulations.

**Table 1** Material properties of copper (Cu) [56], iron (Fe)[57] and their corresponding oxides [58-63].

| Material (Metal) properties | Unit | Cu | Fe |
|---|---|---|---|
| Density | $\mathrm{kg/m^3}$ | 8960 | 7890 |
| Specific heat | $\mathrm{J/kg}-K$ | 383 | 452 |
| $T_m$ | K | 1356 | 1538 |
| Young's modulus | GPa | 124 | 207 |
| Poison's ratio | $-$ | 0.34 | 0.29 |
| $A$ | MPa | 90 | 175 |
| $B$ | MPa | 292 | 308 |
| $n$ | $-$ | 0.31 | 0.32 |
| $C$ | $-$ | 0.025 | 0.06 |
| $m$ | $-$ | 1.09 | 0.55 |
| $T_{ref}$ | K | 298 | 298 |
| Reference strain rate | $1/\mathrm{s}$ | 1 | 1 |

| Material (Oxide) properties | Unit | Cu Oxide | Fe Oxide |
|---|---|---|---|
| Density | $\mathrm{kg/m^3}$ | 6000 | 5240 |
| Young's modulus | GPa | 126 | 220 |
| Poison's ratio | - | 0.31 | 0.37 |
| Fracture toughness | $\mathrm{MPam^{1/2}}$ | 5 | 2.05 |

### 2.2. Peridynamic model and simulation setup

Within the framework of PD, the material of interest is discretized into discrete points known as material points. These points interact through peridynamic bonds within a defined distance range, denoted as $\delta$. To simulate the deformation of the metal particle and substrate, a nonordinary state-based peridynamic material model is employed. Concurrently, an ordinary state-based constitutive model is used to model the deformation and fracture of oxide layers. In PD, the introduction of material damage involves eliminating interactions among material points. Further details on the implementation of these specific peridynamic models, as well as the rationale behind their selection, can be found in our previous studies [5].

Compared with multiparticle models, simulating single-particle interactions provides a valuable approach for understanding the effects of various feedstock parameters while offering greater computational efficiency because it avoids the complexity caused by the interactions between incoming and deposited particles [11,64,65]. In our previous work [52], we demonstrated the effectiveness of two-dimensional (2D) and three-dimensional (3D) peridynamic simulations in accurately describing metal deformation, oxide evolution, separation, and removal. Taking advantage of the simplicity and robustness of the 2D model and considering the axisymmetric nature of the normal impact process [14,66], we prioritize computational efficiency. Specifically, our simulations employ a 2D plane strain model in which the z-degree of freedom of all material points is constrained [66]. As shown in Fig. 1, the dimensions of the substrate (height $H$ and width $D$) are maintained at approximately five times the particle diameter ($d$) to mitigate potential boundary effects [5,14,66]. Notably, the two-dimensional plane strain model has been previously used to study high strain rate deformation and predict the critical velocity in cold spray processes [14,67,68].

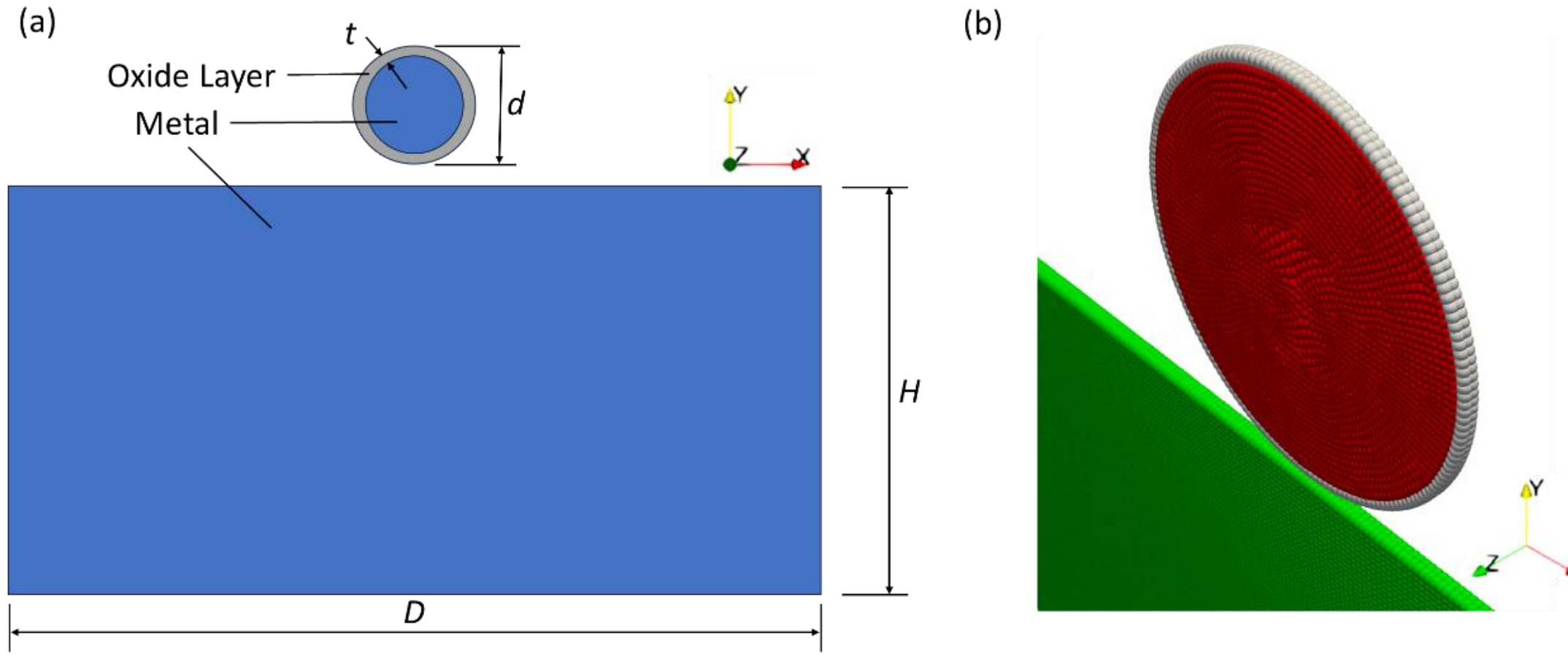

**Fig. 1**. (a) Schematic representation of the PD model used to simulate the CS process, where the particle and substrate materials are colored blue and the oxide layer on the impacting particle is colored gray. (b) The discretization used for the particle, oxide layer and substrate. The green dots represent the substrate, the white dots represent oxides, and the red dots represent metal particles.

### 2.3. Oxide damage model

This study utilizes the critical stretch criterion to simulate the damage behavior of oxide layers. It is assumed that damage will occur when the bond stretch between two material points exceeds a critical value. The bond stretch is defined as follows [69]:

$$s = \frac{|\boldsymbol{\eta}| - |\boldsymbol{\xi}|}{|\boldsymbol{\xi}|} \tag{3}$$

The bond stretch is calculated as the difference between the current bond length $|\boldsymbol{\eta}|$ and the reference bond length $|\boldsymbol{\xi}|$ divided by the reference bond length $|\boldsymbol{\xi}|$. According to the critical stretch failure criterion, a peridynamic bond experiences irreversible breakage when its stretching exceeds a critical value denoted as $s_0$. This critical stretch value is determined by the strain energy release rate $G_0$, which can be experimentally quantified [54,70]

$$s_0 = \sqrt{\frac{G_0}{\left(\frac{6}{\pi}\mu + \left(\frac{16}{9\pi^2}\right)(\kappa - 2\mu)\right)\delta}} \tag{4}$$

Following bond failure, the inability to withstand loads results in the transfer of force to adjacent bonds, leading to localized softening of the material response. This softening promotes the coalescence of broken bonds, leading to damage. With sustained loading, fractures may propagate through the material body [43].

## 3. Results and discussion

### 3.1. Effect of the oxide layer thickness

In the present study, we performed various simulations to evaluate the qualitative and quantitative effects of surface oxide film thickness on particle deformation and oxide removal. In this section, and Section 3.2 below, we focus on Cu as a representative material system, while later, we present the results for the Fe material system.

Figs. 2 and 3 depict the results for plastic strain and temperature evolution, respectively, simulated with oxide films with thicknesses of $t$ = 2.5, 5, 10, 30, and 60 nm on the surfaces of 10 μm diameter Cu particles impacting a Cu substrate. The analysis reveals that the thickness value $t$ affects the deformed particle shape. Specifically, at impact velocities of 450 m/s and 550 m/s, particle deformation is increasingly restrained as the oxide thickness increases. Additionally, at an impact velocity of 650 m/s, where material jetting is evident for $t < 30$ nm, the metal jet is constrained at $t = 30$ nm compared with smaller $t$, and jetting is rendered nonexistent at $t = 60$ nm, as illustrated in Fig. 2 (l) and (o) and Fig. 3 (l) and (o). Earlier experimental investigations involving Cu particles encapsulated in a substantial surface oxide layer (30 nm) impacting a Cu substrate suggested a potential oxide constraint against local deformation on the particle side [71,72]. In the present study, for the first time, we confirm such a constraint on particle deformation via the PD framework.

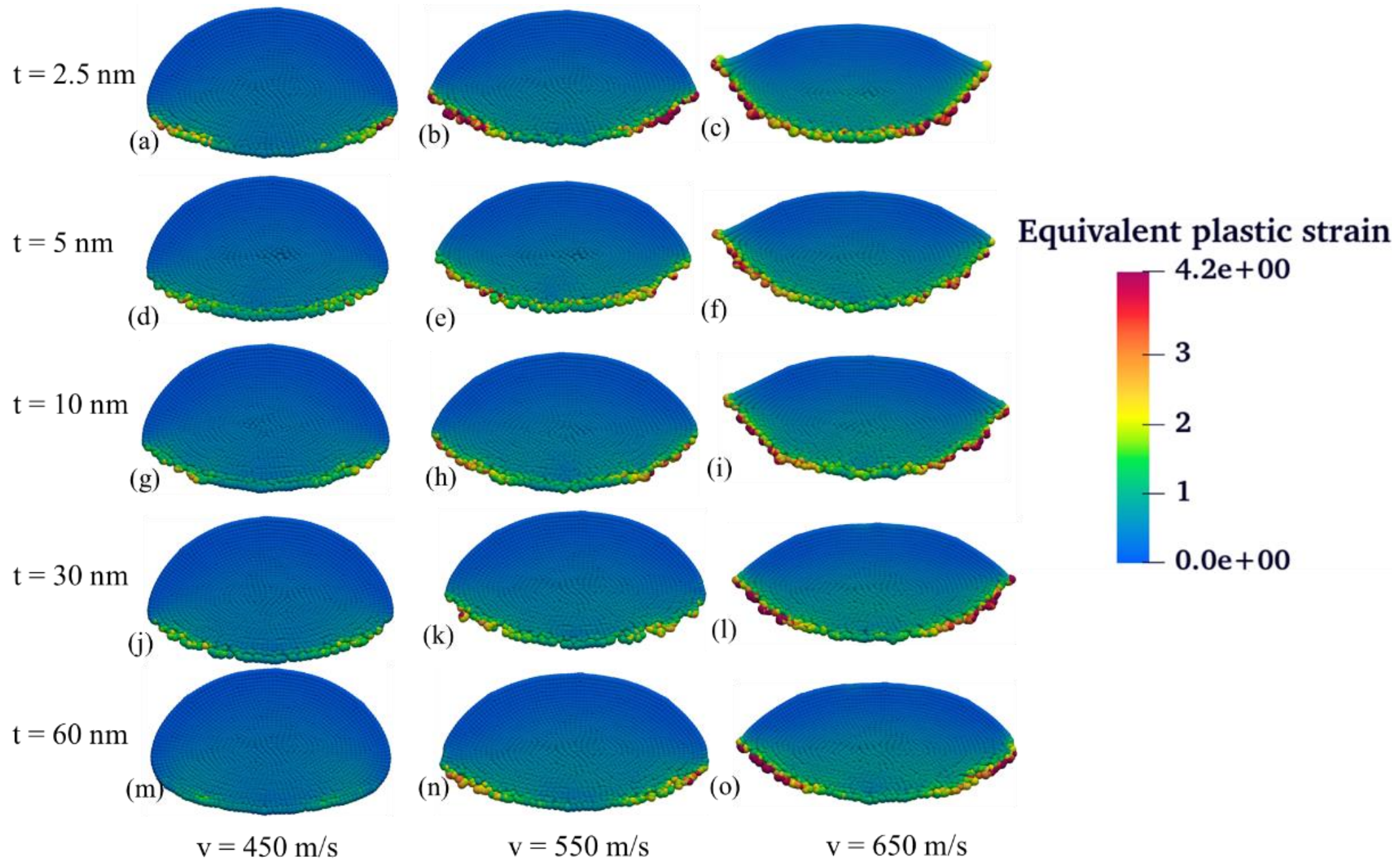

**Fig. 2** Effects of surface oxide film thickness on the evolution of particle plastic strain, simulated at impact velocities of 450, 550, and 650 m/s. These simulations are conducted with Cu particles impacting Cu substrates, i.e., Cu-on-Cu impacts, where the metal particle diameter is $d$ = 10 μm. The color map indicates the magnitude of plastic strain.

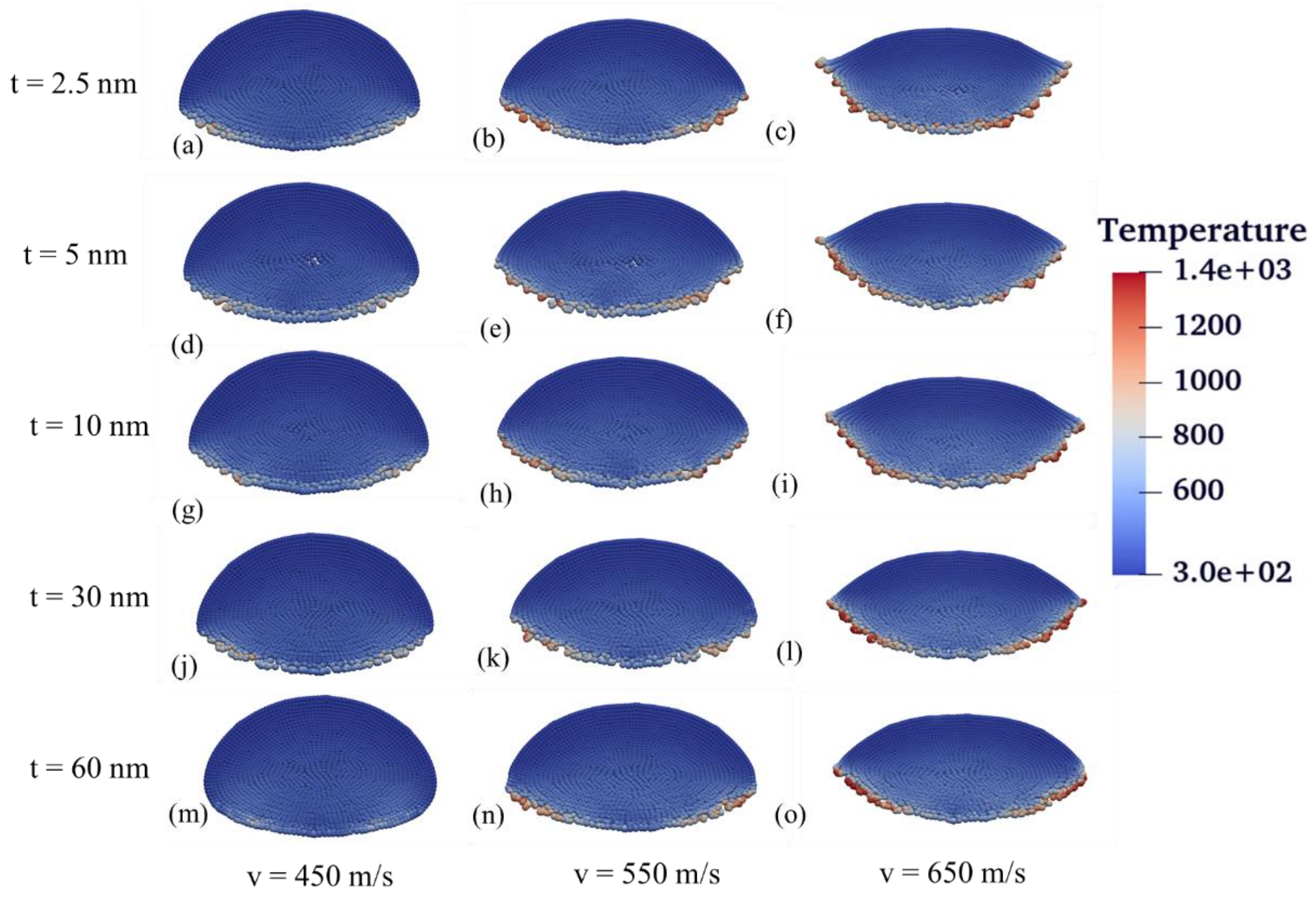

**Fig. 3** Effects of surface oxide film thickness on the evolution of particle temperature, simulated at impact velocities of 450, 550, and 650 m/s. These simulations are conducted with Cu-on-Cu impacts, where the metal particle diameter is $d$ = 10 μm. The color map indicates the magnitude of the temperature.

To provide a quantitative perspective in addition to Figs. 2 and 3 on the effect of oxide thickness on particle deformation, Fig. 4 shows the maximum equivalent plastic strain (MEPS) and maximum temperature (MT) at different oxide thicknesses. This finding indicates that although the MEPS and MT decrease with increasing oxide thickness at lower velocities, the importance of the oxide film on the MEPS and MT decreases at higher impact velocities. Specifically, at an impact velocity of 550 m/s, the MEPS and MT only exhibit a reduction from an oxide thickness of 30 nm to 60 nm. Moreover, at an impact velocity of 650 m/s, there is a negligible change in both the MEPS and MT. The disparate trends observed for the MEPS and MT at various velocities may be attributed to the varying thicknesses of the oxide layer. Fig. 5 shows the final oxide distributions for particles with outer oxide layers of different thicknesses at three impact velocities: 450, 550, and 650 m/s. In all instances, sufficiently high velocities lead to the disruption and ejection of the oxide layer on the particle's bottom surface, facilitating intimate contact of freshly exposed metallic surfaces. However, these effects are considerably impeded at relatively low impact velocities (i.e., 450 m/s) as the thickness of the oxide layers increases. At this velocity, the entire oxide layer at the bottom of the particle with a 60 nm oxide layer remains intact (Fig. 5 (m)), thereby limiting particle deformation. Additional insights are provided by the damage parameter contours within the oxide layer at the impact instances shown in Fig. 6. The constitutive model employed to characterize the brittle behavior of the oxide layer effectively describes its failure during impact. When considering a 60 nm oxide layer around a 10 μm diameter particle, a low velocity of 450 m/s fails to fully damage the oxide layer at the periphery where the most severe deformation is expected, in contrast to the oxide layers of the same particle, which are thinner. At higher velocities, the oxide layer is disrupted on the particle's bottom surface for all the considered oxide thicknesses, allowing for more substantial deformation. Consequently, the MEPS and MT remain nearly unchanged at a velocity of 650 m/s, as shown in Fig. 4.

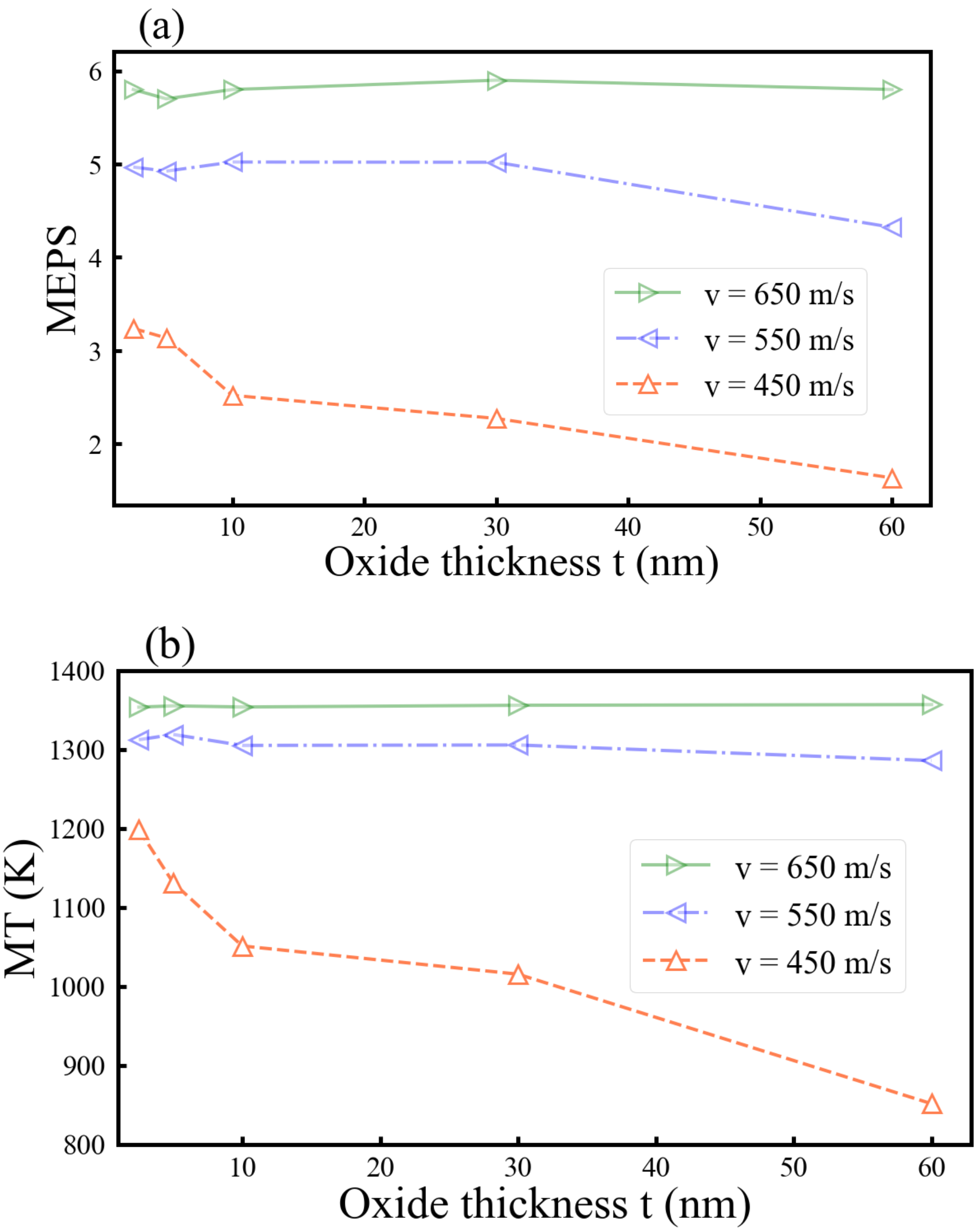

**Fig. 4** Effects of surface oxide film thickness *t* on the particle (a) maximum equivalent plastic strain (MEPS) and (b) maximum temperature (MT) simulated with impact velocities of 450, 550, and 650 m/s. The simulations are conducted with Cu-on-Cu impacts, where the metal particle diameter is *d* = 10 μm.

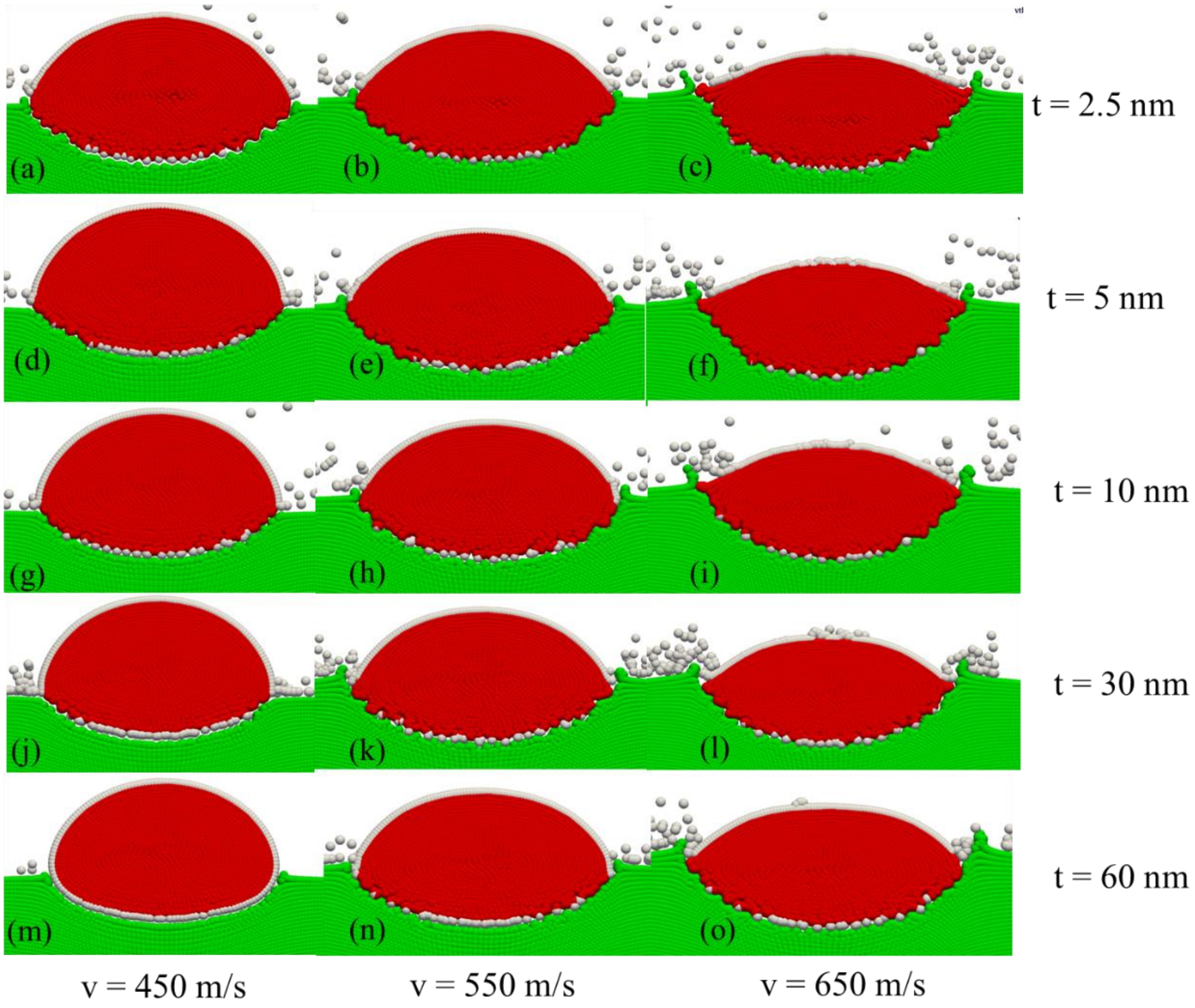

**Fig. 5** Effects of surface oxide film thickness on the particle oxide distribution simulated with impact velocities of 450, 550, and 650 m/s. These simulations are conducted with Cu-on-Cu impacts, where the metal particle diameter is $d$ = 10 μm. In the figure, green dots represent the substrate, white dots represent oxides, and red dots represent particles.

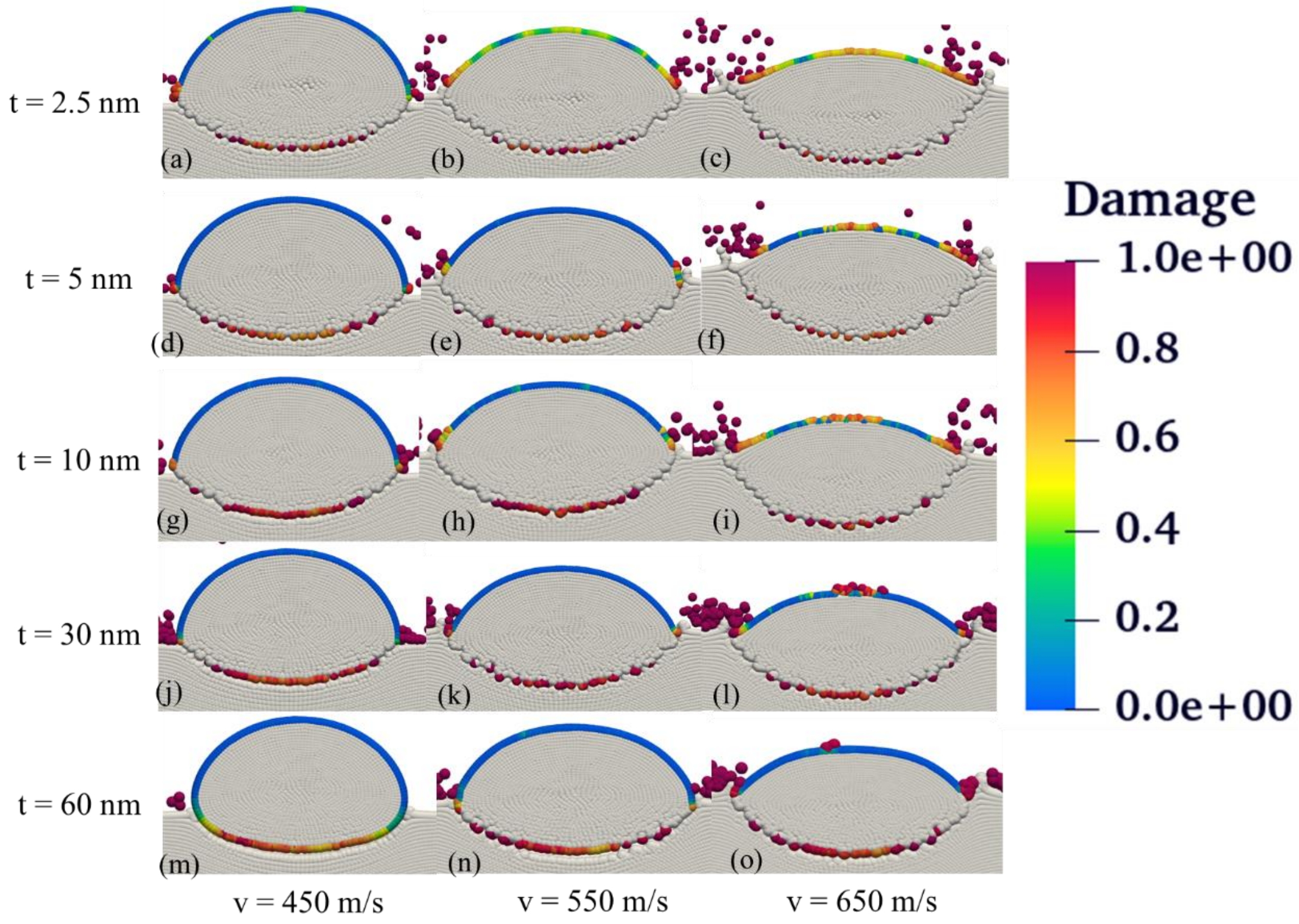

**Fig. 6** Effects of surface oxide film thickness on particle oxide damage simulated with impact velocities of 450, 550, and 650 m/s. The simulations are conducted with Cu-on-Cu impacts, where the metal particle diameter is $d$ = 10 μm. The visualization uses blue and red colors to represent the extent of damage to oxides, whereas white colors represent particles and substrates.

An intriguing observation in Fig. 5 and Fig. 6 is the initiation of jetting in cases with varying oxide layer thicknesses. Specifically, at an impact velocity of 450 m/s, the particle covered by a 2.5 nm oxide layer induces material jetting from the substrate at the contact edge (Fig. 5 (a)), whereas such jetting is not observed in cases with thicker oxide layers. On the other hand, particles at a velocity of 650 m/s exhibit jetting phenomena, except for those with a 60 nm oxide layer (Fig. 5 (o)). The initiation of material jetting is a key indicator of contact behavior and bond strength during CS deposition, as observed in experimental results [72] and our previous numerical studies [52]. Two cases are distinguished: one where jetting occurs only on the substrate, below the critical velocity, indicating limited metallurgical bonding due to insufficient oxide removal; and another where jetting occurs on both the substrate and particles, exceeding the critical velocity, indicating the removal of oxides in large quantities and resulting in the formation of high-strength metallic bonds. For the thinnest oxide layer of 2.5 nm, jetting on the substrate begins at 400–450 m/s, and jetting on both the substrate and the particle starts between 550–600 m/s. In contrast, for the thickest oxide layer of 60 nm considered in this study, substrate jetting is not observed until 550–600 m/s, whereas particle jetting occurs at impact velocities between 700–750 m/s. The simulation results for oxide thicknesses of 2.5 nm and 60 nm at the velocities where jetting is first observed are depicted in Fig. 7. For oxide thicknesses ranging from 5 nm to 30 nm, our findings indicate the onset of substrate-only jetting at velocities between 500–550 m/s, and at 600–650 m/s, jetting is observed on both the substrate and the particle. As summarized in a previous study [73], for a 10 μm copper particle with a 2.5 nm oxide layer thickness, the critical velocity was determined to be 470 m/s. The initiation of substrate jetting at an impact velocity of 400–450 m/s, as obtained in the current study, appears reasonable. The critical velocity for a 10 μm particle with a thick oxide layer of 60 nm can be estimated to be in the range of 650–700 m/s, which requires further validation through experiments. Fig. 8 shows the effect of the oxide layer thickness on the residual oxide at the interface after impact, revealing that the thicker the oxide layer is, the larger the volume of residual oxide, thereby increasing the critical velocity and adversely affecting deposition. These findings are consistent with experimental studies [73]. The substantial decrease in velocities necessary for initiating jetting from an oxide thickness of 5 nm to 2.5 nm corresponds well with previous experimental results when considering copper oxide thicknesses ranging from 20 nm to 2 nm. This alignment indicates heightened sensitivity of the critical velocity to the thinning oxide layer [73]. Furthermore, our simulations indicate a significant elevation in critical velocity from an oxide thickness of 30 nm to 60 nm, emphasizing the need for additional experimental validation.

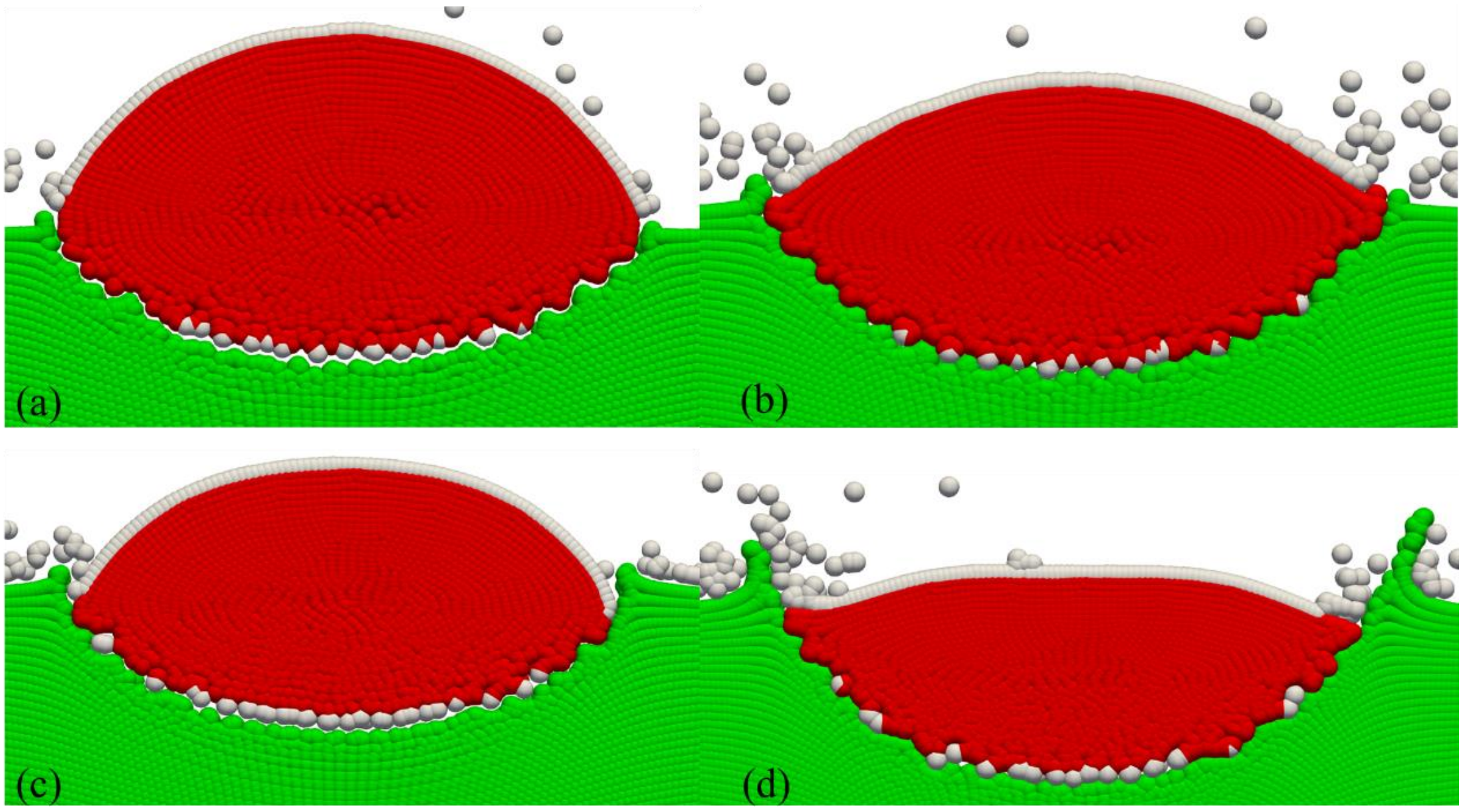

**Fig. 7** Close-up views showing the effect of oxide layer thickness on the initiation of jetting from the substrate and/or particle, illustrated by a few sample cases of different oxide thicknesses and particle velocities: (a) $t$ = 2.5 nm, $v$ = 450 m/s; (b) $t$ = 2.5 nm, $v$ = 600 m/s; (c) $t$ = 60 nm, $v$ = 600 m/s; (d) $t$ = 60 nm, $v$ = 750 m/s. These simulations are conducted with Cu-on-Cu impacts, where the metal particle diameter is $d$ = 10 μm. In the figure, green dots represent the substrate, white dots represent oxides, and red dots represent particles.

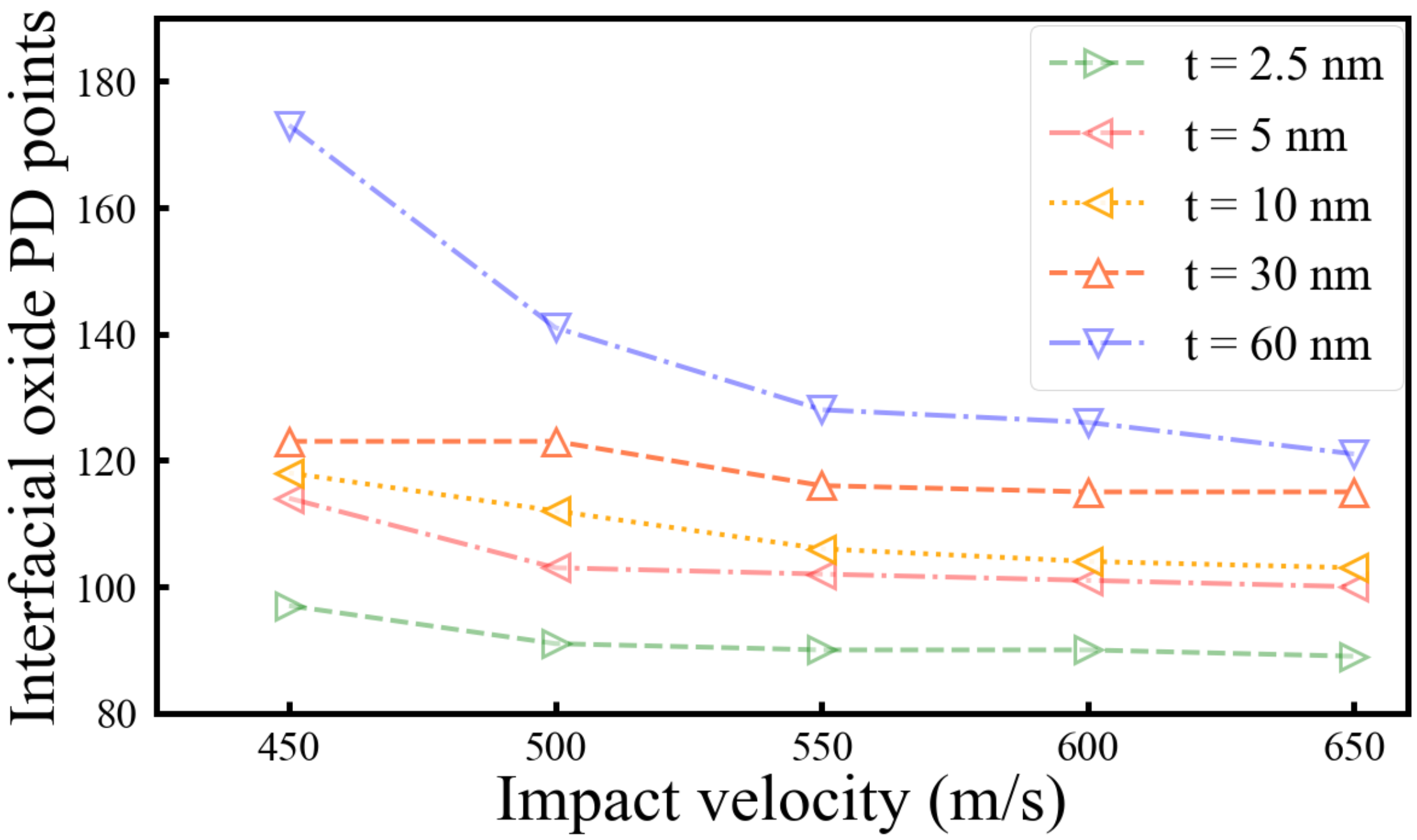

**Fig. 8** Effect of the oxide layer thickness *t* on the remaining oxide at the interface after impact. The simulations are conducted with Cu-on-Cu impacts, where the metal particle diameter is $d$ = 10 μm.

Previous investigations have shown that the residual oxide layer at the interface can hinder the adhesion strength and reduce the extent of surface bonding [28,74]. Our simulations demonstrate the fracture of the oxide layer into debris at the contact interfaces, with the oxide films on the particle surfaces undergoing partial deposition into the coating, potentially influencing the microstructure and performance of the coating. Notably, an increase in oxide thickness correlates with an increased residual oxide layer at the interface, as illustrated in Fig. 8. This implies that a thicker oxide layer not only elevates the critical velocity but also compromises the quality of the coating.

### 3.2 Effect of particle size

The size effect plays a significant role in practical scenarios in CS applications where the particle size may vary widely, spanning more than an order of magnitude [75]. This variability in particle size is a crucial factor influencing various aspects of the process, such as the particle velocity and temperature and bonding in the coating [76]. Schmidt *et al.* [13] and Dowding *et al.* [77] conducted studies that highlighted the particle size effect in CS, specifically in materials such as Cu, 316 L stainless steel, pure Al, and Ti particles impacting matching substrates. Their experimental results indicated that an increase in the average particle diameter could lead to a lower critical velocity for the same materials in the CS. Computational methods have also been employed to explore the effect of particle size on critical velocities [11,13,33,77,78]. These studies align with experimental findings, suggesting that finer particles tend to exhibit higher critical velocities. However, it is crucial to note that previous numerical simulation results might lack reliability because of the potential influence of the oxide layer, which many studies did not consider. In cases where the oxide layer was indeed incorporated into models, its thickness was often several hundreds of nanometers, which is significantly greater than the actual oxide thickness observed in experiments. Addressing this discrepancy is essential for more accurate simulations and a better understanding of the size effect in CS applications.

This section focuses on the impact of Cu particles on Cu substrates with a 10 nm oxide layer. Fig. 9, which illustrates the deformed shapes of the particles ranging from 5--50 μm in size at various impact velocities, reveals a noteworthy observation: the overall deformed shapes of the particles appear similar regardless of their size at the same impact velocity. This finding suggests uniformity in deformation across different particle sizes, which aligns with observations from previous FEM simulations [79]. Further insights are provided in Fig. 10, which shows the variation in the compression ratios with the particle impact velocity for various particle sizes. The compression ratios clearly exhibit similarities among different particle sizes when subjected to the same impact velocity.

An essential indicator during the particle impact process is the initiation of discontinuities within the oxide, indicating complete damage. Our prior study [52] and experiments in the literature [71,72] revealed that this onset marks the exposure of fresh metal-to-metal contact, potentially influencing the rebound behavior of particles. As depicted in Fig. 11, this initiation occurs at 400 m/s for the 10 μm copper particles, whereas the 5 μm particles require an impact velocity of 500 m/s. This implies that larger particle diameters necessitate lower velocities for inducing bonding. Fig. 11 also illustrates the initiation of jetting on both the substrate and the particle, along with the corresponding impact velocities for the 5 μm and 10 μm copper particles. As discussed earlier, permanent bonding occurs within the velocity range between these two jetting onsets. For the 5 μm particle, the critical velocity exceeds 600 m/s, requiring an impact velocity of 700 m/s to achieve robust bonding, as indicated by the onset of jetting on both the substrate and the particle. Conversely, the 10 μm copper particle exhibited a corresponding velocity range of 500–650 m/s, suggesting a lower critical velocity than the 5 μm particle.

As anticipated from previous studies [77], the expectation that larger particles, endowed with higher kinetic energy, necessitate lower velocities to initiate jetting is confirmed. This paper uniquely underscores the explicit modeling of nanoscale oxide layers covering the particles, revealing novel insight. This signifies, for the first time, that larger particles not only demand lower velocities for material jet initiation but also for inducing damage and separation of the oxides, thus lowering the velocity for initiating the formation of metallurgical bonds.

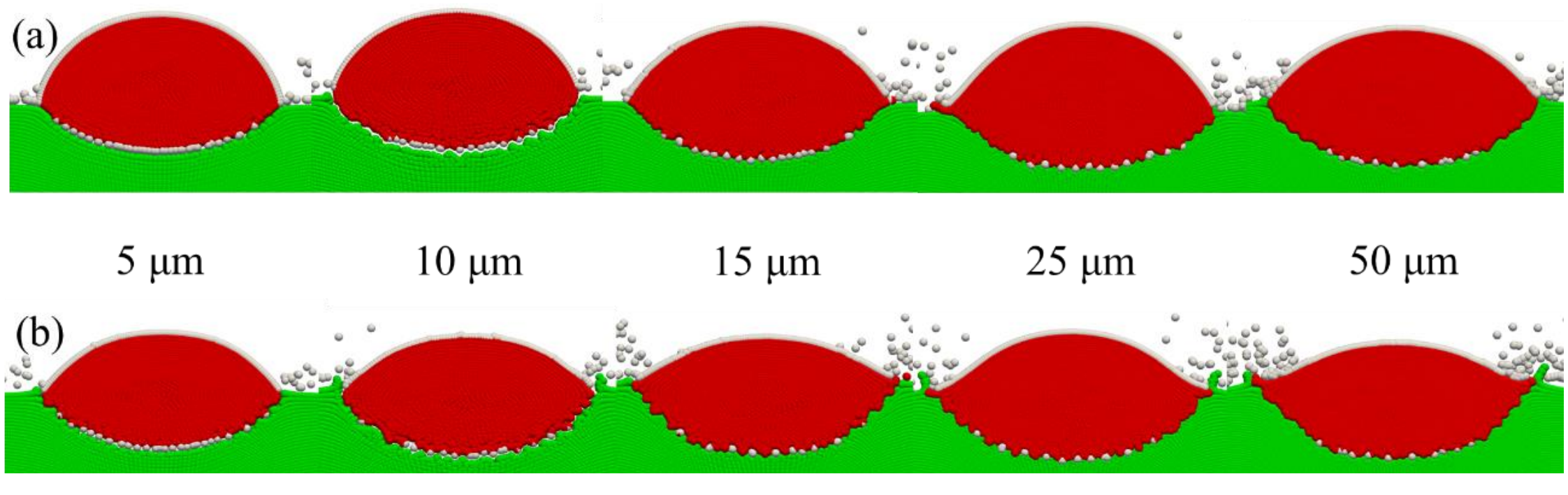

**Fig. 9** Effects of particle size on the remaining oxide at the interface after Cu-on-Cu impact at particle velocities of (a) 500 m/s and (b) 600 m/s. The metal particle diameters range from $d$ = 5 μm to 50 μm. The metal oxide is 10 nm thick. In the figure, green dots represent the substrate, white dots represent oxides, and red dots represent particles.

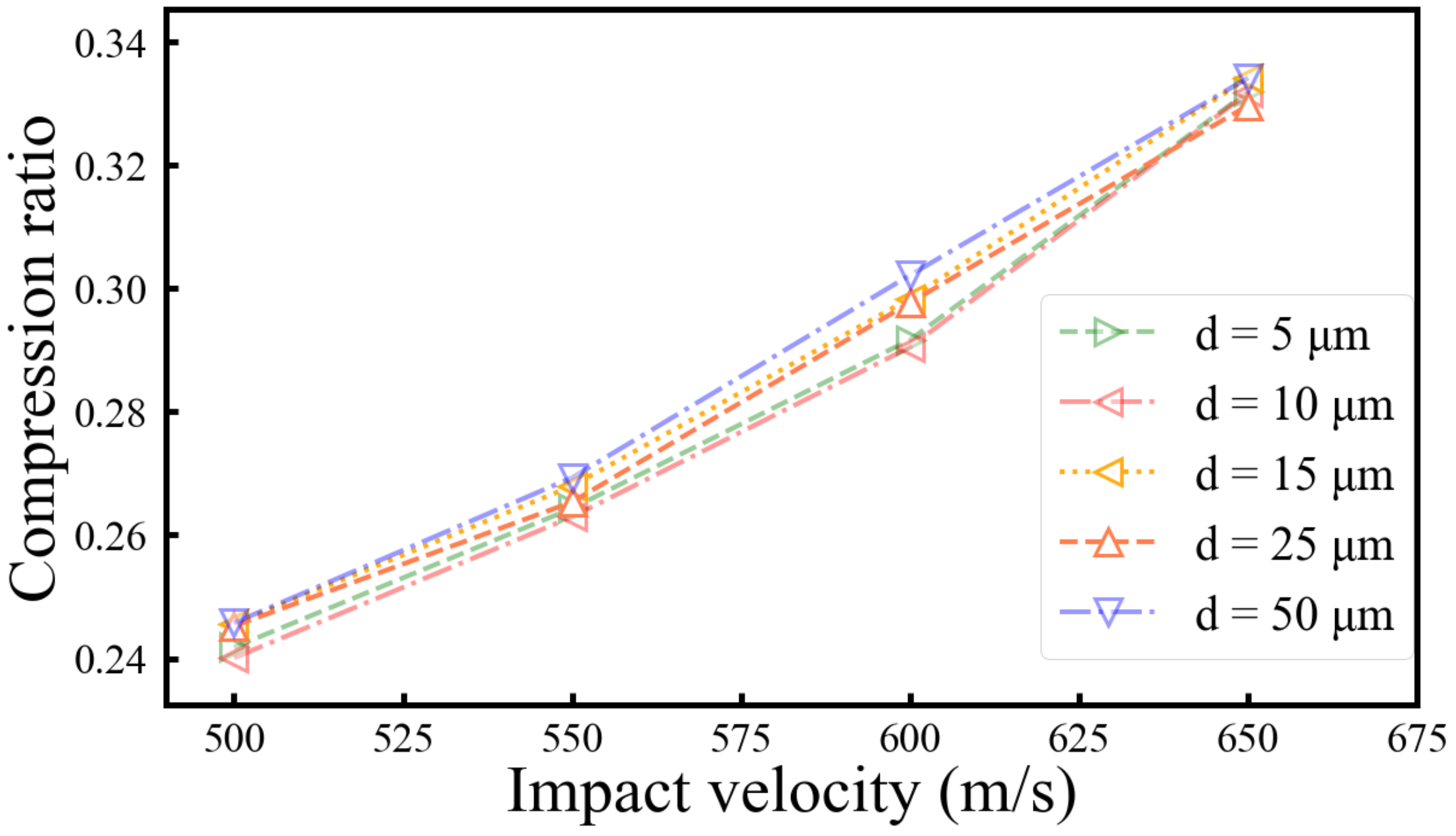

**Fig. 10** Effect of particle size on the compression ratio for Cu-on-Cu impact. The metal oxide is $t$ = 10 nm thick.

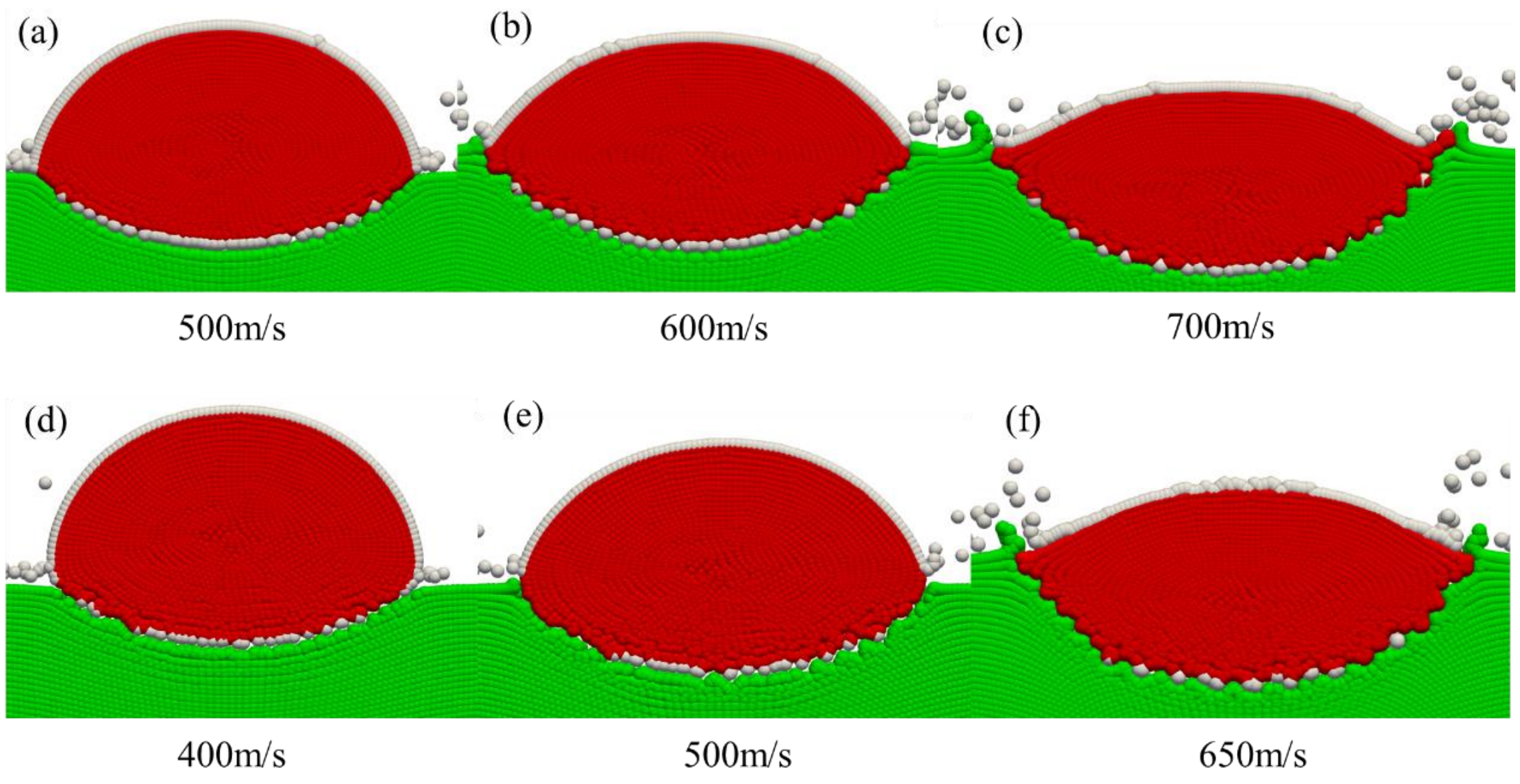

**Fig. 11** Initiation of oxide discontinuity and jetting for Cu-on-Cu impact. Panels (a)-(c) depict particles with diameters of 5 μm, while panels (d)-(f) show particles with diameters of 10 μm, each covering a 10 nm oxide layer. In the figure, green dots represent the substrate, white dots represent oxides, and red dots represent particles.

## 3.3 Effects of the particle material

The results in the previous sections are based on a Cu particle impacting a Cu substrate. To gain a more general understanding, we also consider the case in which the material is Fe. Fig. 12 shows some sample side–to–side comparisons in terms of the final deformation and oxide distribution profiles obtained at varying velocities for the two material systems. The results indicate that although the general phenomenon of oxide film breakage occurring at high velocities remains the same, there are some notable differences between the two material systems. For the Fe-on-Fe impact, the oxide film at the contact region is fully cracked when the impact velocity is greater than 550 m/s, whereas for the Cu-on-Cu impact, the oxide film cracks at a lower velocity ranging from 450–500 m/s. Moreover, for the Fe-on-Fe impact, despite the high-velocity impact breaking up the oxide film, a more significant portion of cracked oxide persists at the interface because of the limited deformation compared with that of the Cu-on-Cu case.

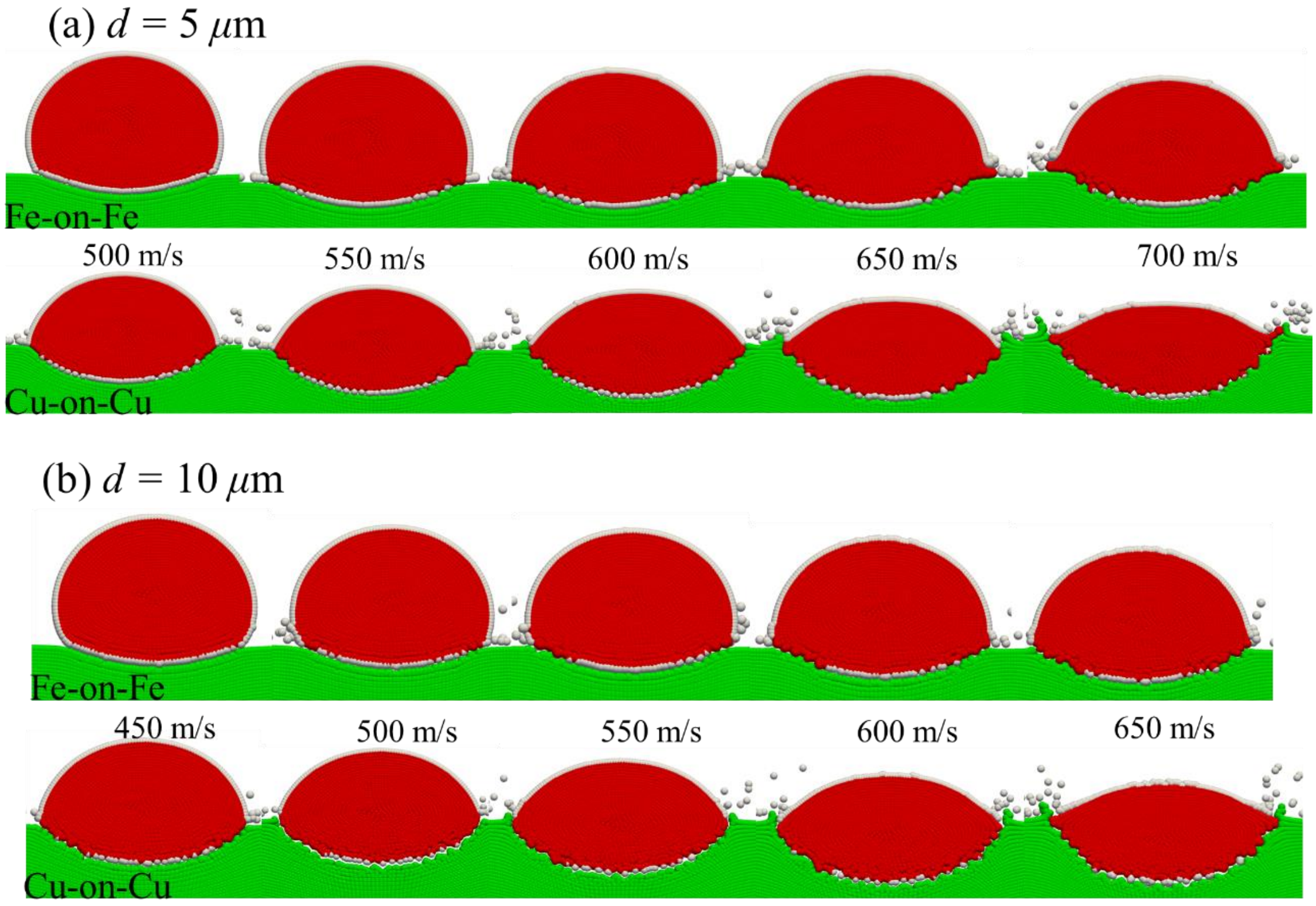

**Fig. 12** Effects of material type on particle size: a) $d = 5\ \mu$m b) $d = 10\ \mu$m, with 10 nm oxide. In the figure, green dots represent the substrate, white dots represent oxides, and red dots represent particles.

Increasing the impact velocity to 600 m/s reveals material jets at the substrate in the two Cu–

Cu cases, i.e., the particle sizes $d = 5\mu m$ and $10\mu m$, aiding the expulsion of cracked oxides from the interface center. While some oxides remain at the interface, their quantity is smaller than that in the Fe-on-Fe cases. Additionally, it is clear from the final deformation profiles of the particle and substrate that at the same impact velocity, the Cu–Cu interface has a larger contact area than the Fe–Fe interface does, which is attributed to more extensive deformation in the Cu–Cu impact. This observation can be further quantitatively supported by plotting the compression ratios for the two material systems, as shown in Fig. 13(a), where higher compression ratios are observed for the Cu-on-Cu impact. Additionally, the amount of oxide residue after impact, as shown in Fig. 13(b), demonstrates a greater proportion of removed oxide for the Cu-on-Cu impact. These findings suggest that, at the same impact velocity, it would be easier for soft particles against a soft substrate (e.g., Cu-on-Cu) to establish metallic bonding because of the larger interface area and, as shown in Fig. 13(b), greater oxide removal than hard particles against a hard substrate (e.g., Fe-on-Fe). Notably, for the Cu–Cu combination, the material jet formed at the rim of the substrate crater, whereas for Fe–Fe, the substrate crater lacked a metal jet. In general, the material jet at the substrate crater rim contributes partially to removing cracked oxides, further enhancing the bonding strength of the Cu–Cu interface.

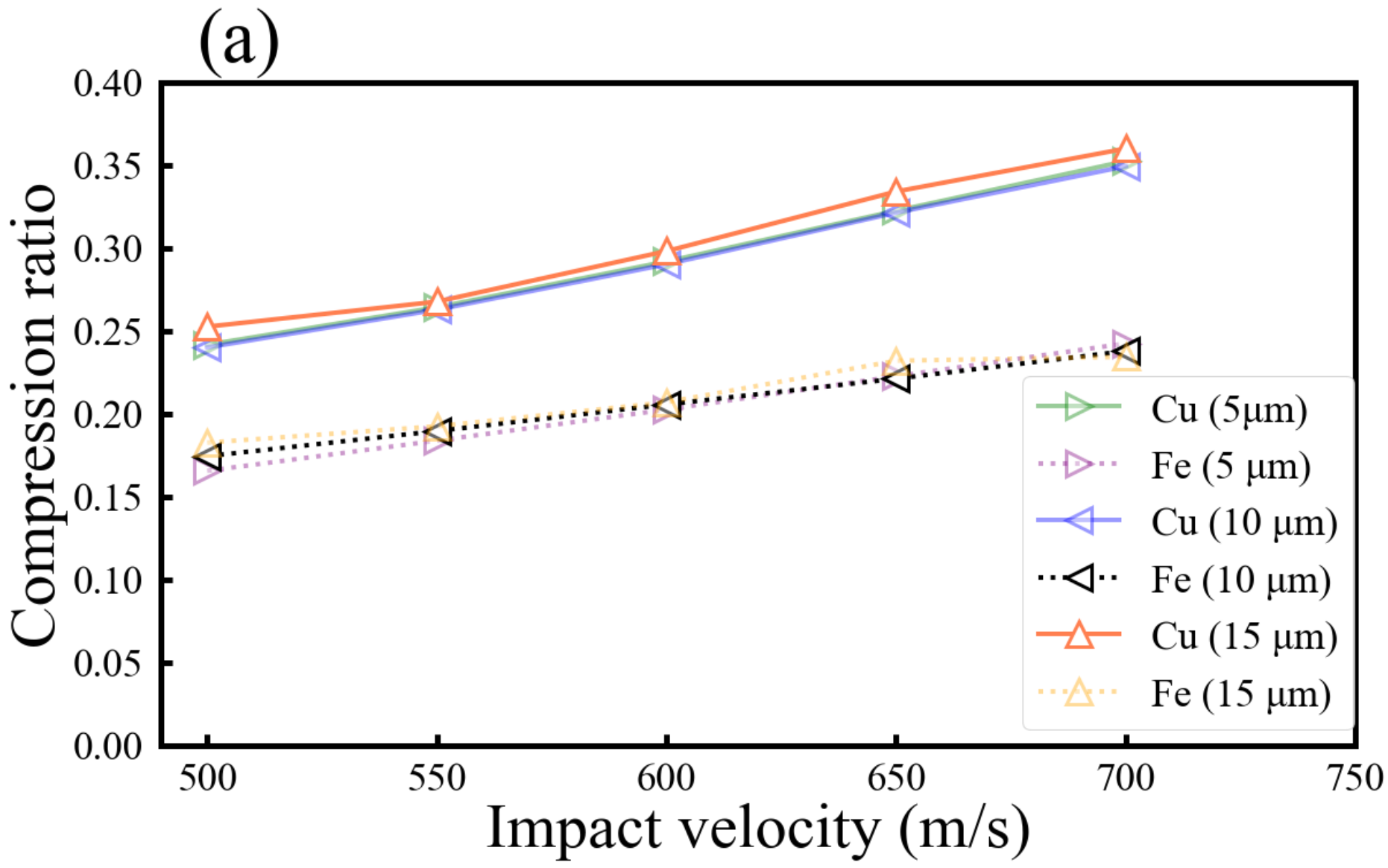

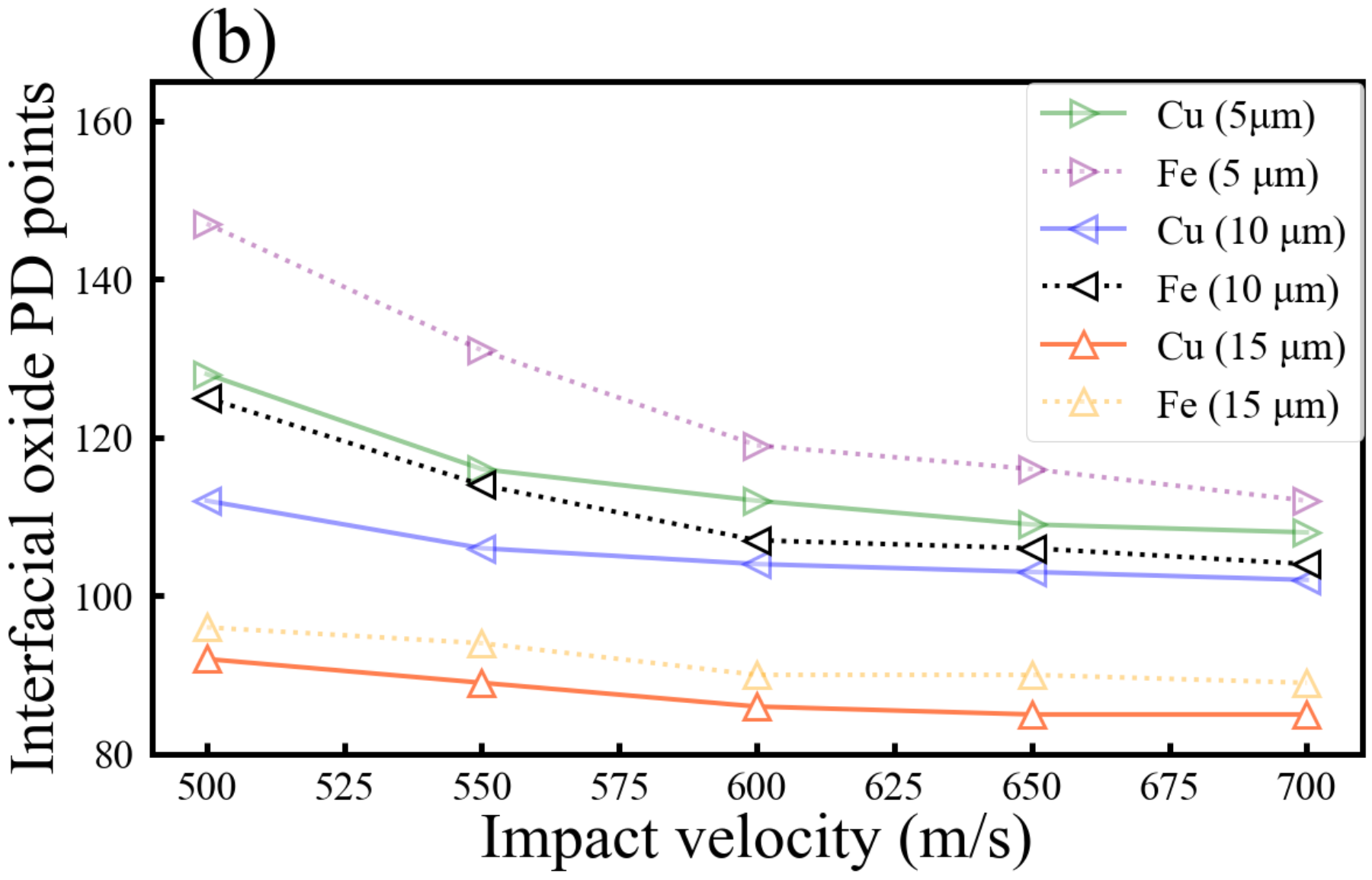

**Fig. 13** (a) Compression ratio and (b) amount of trapped oxide (namely, the amount of interfacial oxide $N_I$) of Cu-on-Cu impacts and Fe-on-Fe impacts as the impact velocity varies from 500–700 m/s. The results shown here are for three representative particle sizes, namely, 5 μm, 10 μm and 15 μm, all with 10 nm oxides.

Additionally, we can see from Fig. 13(b) that as the velocity increases to 650 m/s and 700 m/s, despite more pronounced deformation of the particle and substrate, the amount of remaining oxide at the interface appears to change little in both the Cu-on-Cu and the Fe-on-Fe cases, indicating that the quantity of remaining oxide is insensitive to the impact velocity at very high velocities (i.e., from 650 m/s). Another observation from Fig. 13(b) is that the residual oxide for both Cu-on-Cu and Fe-on-Fe decreases with increasing particle size, and we argue that this may represent some characteristics of the bonding phenomenon in CS, which is further discussed in the next section.

### 3.4 Effect of the particle size on the critical velocity

The results presented above clearly indicate that the particle size affects the material deformation and oxide breakage/removal processes and that, naturally, the particle size affects the critical velocity. Experimental observations [13,77] have indicated that the critical velocity, denoted as $V_c$, decreases as the particle size increases, a relationship that can be mathematically described by a power law equation:

$$V_c \propto D^i \tag{5}$$

Previous studies [13,77] determined that the size scaling index, denoted as $i$, is approximately $-0.2$. In particular, for Cu, $i = -0.19$; for 316 L stainless steel, $i = -0.14$; for Al, $i = -0.19$; and for Ti, $i = -0.21$. In a related development, Hassani-Gangaraj *et al.* [80] formulated an approximation for the critical velocity that relies on a hydrodynamic spall process:

$$\frac{V_c}{C_0} \approx \frac{2}{k} \times \frac{P_s}{B} \tag{6}$$

where $C_0$ represents the shock velocity, $B$ is the bulk modulus, and $P_s$ denotes the spall strength. Eq. (4) is founded on the dynamics of the shock wave emerging upon contact. When this shock wave reaches the particle's edge, it can initiate a release wave, leading to jetting if the stresses surpass the local spall strength of the material [80]. Building on the groundwork of Hassani-Gangaraj *et al.* [80], Dowding *et al.* [77] proposed a mechanism for explaining the particle size dependence via the FEM. The authors proposed that adiabatic heating induced softening at the interface, thereby lowering the spall strength and reducing the barrier for jetting and bonding [77]. Larger particles, endowed with higher kinetic energy, demand a lower impact velocity to initiate jetting. Their simulations yielded a satisfactory prediction for the impact bonding of Al and Ti, which aligns well with experimental observations. Notably, the authors of [21] did not consider oxide layers in their numerical models.

Indeed, impact-induced bonding for supersonic particles has been proposed to be closely linked to jetting phenomena [14,23,73,80,81]. While the origin of jetting in the realm of CS remains a topic of debate, models grounded in adiabatic shear instability [2,11,45] and hydrostatic pressure spall [51,52] have been posited to explain the onset of jetting. Jetting disrupts surface oxide films, regardless of the specific mechanism behind their formation. This disruption is considered a mechanism for cleaning the surface, facilitating pristine contact between the particle and substrate, and ultimately leading to bonding in the CS [14,82]. Consequently, this section focuses on the concept that, with increasing particle diameter, the residual oxide volume decreases at the same impact velocity—an inference derived from Fig. 13(b). This implies that larger particles require smaller impact velocities to achieve bonding. To quantitively investigate the effect of particle size, we initially conducted PD simulations akin to those depicted in Fig. 12. These simulations encompassed a variety of particle sizes for both Cu-on-Cu and Fe-on-Fe impacts, with a specific focus on quantifying the amount of oxide situated at the particle/substrate interface, as illustrated in Fig. 13(b). To address potential discrepancies arising from discretization variances inherent in the PD models across the entire particle size range, we applied the following normalization to the raw data:

$$N = \frac{N_I}{N_T} \tag{7}$$

where $N_I$ represents the number of interfacial oxide PD points and where $N_T$ is the total number of oxide material points covering the entire metal particle. This normalization is considered meaningful, as it offsets the fluctuation of $N_I$ across various particle sizes, since $N_I$ is expected to increase as the particle size increases, reflecting the enlargement of the contact region with no or little oxide removal. The parameter $N$ thus represents the portion of the total oxide encapsulated within the interface. Fig. 14 shows that for a particular particle size, $N$ varies with impact velocity, increasing monotonically as the velocity increases prior to the onset of oxide discontinuity, which aligns with our expectation. On the other hand, $N$ decreases with increasing velocity after the onset of discontinuity, indicating the onset of the oxide cleaning effect. Furthermore, in Figs. 15(a) - (d), before the onset of oxide discontinuity (at low impact velocities, e.g., 150 m/s and 300 m/s), the parameter $N$ remains unchanged as the particle size varies for both the Fe and the Cu material systems. Moreover, $N_I$ monotonically increases as the particle size increases, which is in alignment with our expectation. This confirmation establishes $N$ as a more reliable indicator of the extent of damage and oxide removal, since changes in $N$ are solely attributable to the crushing and cleaning effects caused by particle deformation and are independent of particle size.

On the other hand, once the impact velocity is sufficient to induce oxide discontinuity, e.g., at high impact velocities of 450 m/s and 550 m/s, as shown in Fig. 16, $N$ monotonically decreases as the particle size increases. Notably, the selection of impact velocities depicted in Fig. 16 is contingent upon the minimum velocity at which the cleaning effect begins to manifest for the smallest particle size investigated in this study (5 μm). Accordingly, we chose impact velocities of 450 and 500 m/s for the Cu-on-Cu impacts and 550 and 600 m/s for the Fe-on-Fe impacts. In particular, the evolution of $N$ as a function of the particle size exhibits a power-law form. Fitting the data to the power-law form, expressed as $N = N_0 D^{\alpha}$ with $N_0$ and $\alpha$ as fitting parameters, we obtained the values of $N_0$ and $\alpha$ for the Fe-on-Fe and Cu-on-Cu systems, as listed in Table 2.

**Table 2** $N_0$ and $\alpha$ for the Fe-on-Fe and Cu-on-Cu systems

| | $N_0$ | $\alpha$ |
|---|---|---|
| Fe-on-Fe at 550 m/s | 0.3663 | −0.199 |
| Fe-on-Fe at 600 m/s | 0.3327 | −0.192 |
| Cu-on-Cu at 450 m/s | 0.3942 | −0.236 |
| Cu-on-Cu at 500 m/s | 0.3662 | −0.225 |

Table 2 shows that $\alpha$ is more negative for Cu-on-Cu than for Fe-on-Fe. This is consistent with our previous observations that oxide removal is easier in the Cu-on-Cu cases than in the Fe-on-Fe cases (see Fig. 12 and Fig. 13b). As discussed earlier, if oxide removal is considered the critical condition for establishing metallic contact and achieving bonding, then the parameter $\alpha$ would directly correspond with the parameter $i$ in Eq. (5). Table 2 demonstrates that the obtained values of $\alpha$ are consistent with the observations of Schmidt *et al.* [13], who reported $i = -0.14$ for stainless steel and $i = -0.19$ for Cu. Thus, we assert that the proposed mechanism for the size effect encapsulates the fundamental physics: larger particles result in smaller oxide-covered interfaces, thereby requiring lower impact velocities to achieve bonding conditions.

The data presented in Fig. 16 represent, to our knowledge, the first numerical attempt to elucidate the effect of particle size on impact bonding while considering the influence of the oxide layer, which is a crucial factor in the CS process. Notably, this methodology remains relevant in scenarios or materials where particle jetting is not visibly present even at critical velocities. For example, in instances where Cu particles do not exhibit jetting signs at or slightly above the corresponding critical velocity [71,72], the size effect still exists and may be attributed to oxide breakage and removal resulting from plastic deformation.

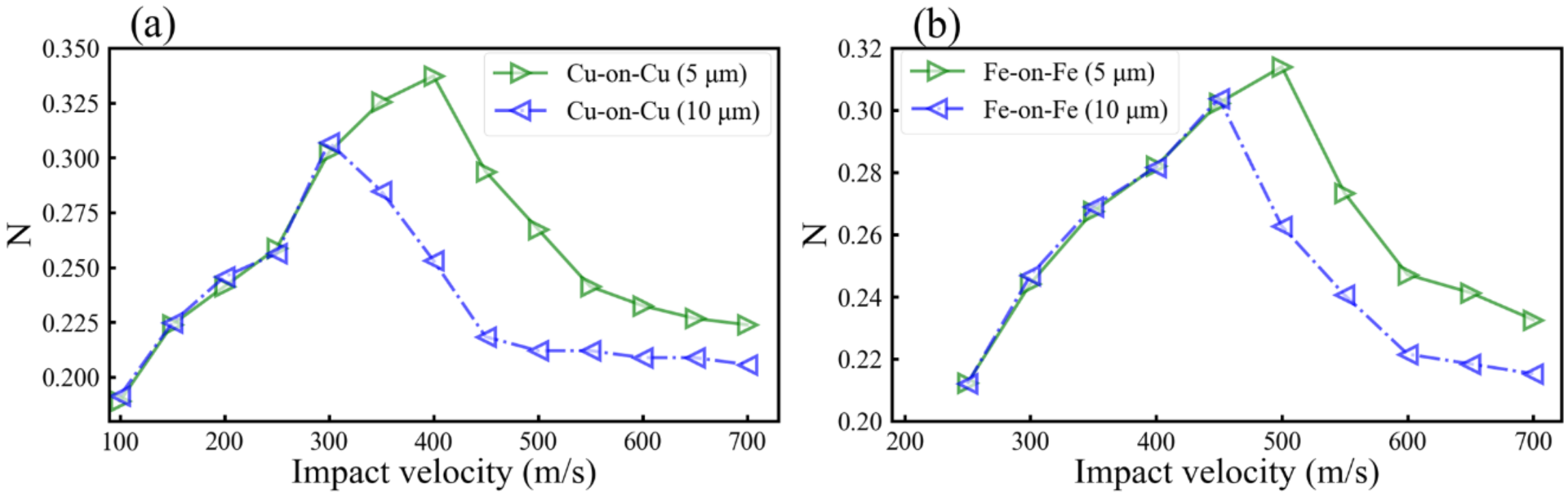

**Fig. 14** Normalized number of trapped oxides $N$ of a) Cu-on-Cu and b) Fe-on-Fe impacts as a function of the particle velocity. The oxide is 10 nm thick.

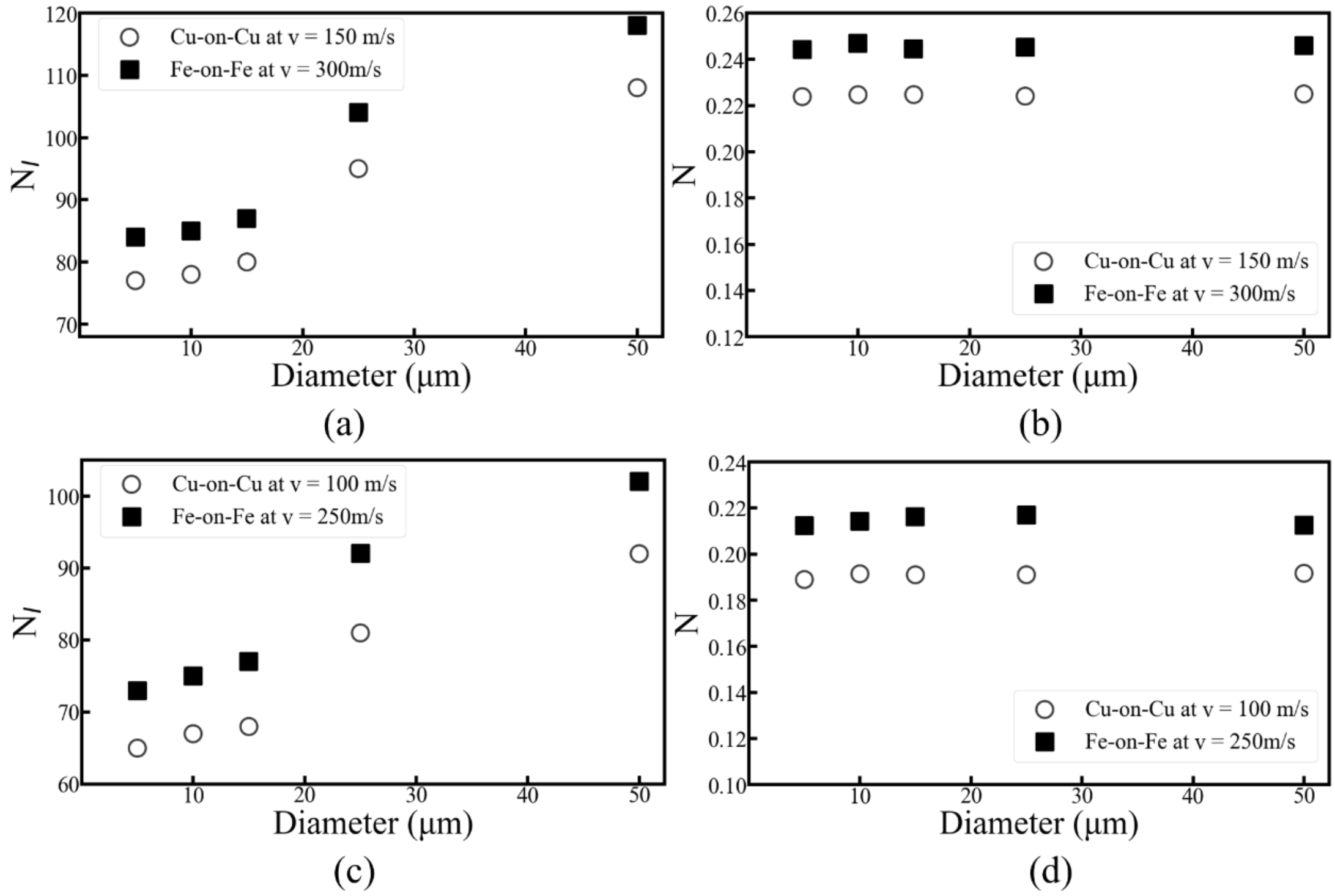

**Fig. 15** Peridynamic results for a) and c) the number of trapped oxides for Cu-on-Cu impacts at 150 m/s and Fe-on-Fe impacts at 300 m/s and b) and d) the normalized number of trapped oxides for Cu-on-Cu impacts at 100 m/s and Fe-on-Fe impacts at 250 m/s as a function of the particle size.

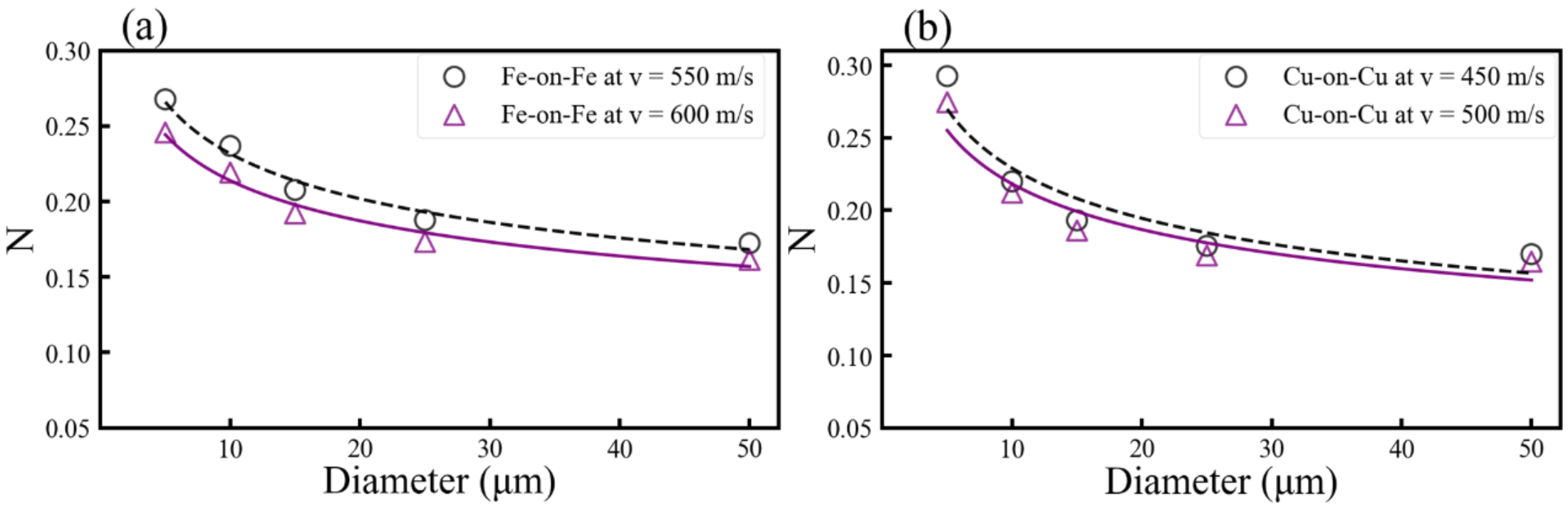

**Fig. 16** Normalized number of trapped oxides $N$ of a) Fe on Fe and b) Cu on Cu as a function of the particle size. The oxide is 10 nm thick.

## 4. Conclusion

By considering an oxide layer around a metal particle, this study systematically investigated the material deformation and oxide fracture process of a single metal particle cold-sprayed to impact a matched metal substrate via peridynamics (PD) simulations. Various aspects, including the oxide thickness, particle/substrate material, and particle size, have been examined in detail. Our results led to several findings, with the main findings summarized below.

(a) A thicker oxide film on Cu particles restricts deformation, affects plastic strain and temperature at the interface, and delays oxide discontinuity and material jetting. This increases the critical velocity required for metal–metal contact, which is consistent with the experimental results.

(b) Different particle sizes exhibit similar compression ratios at the same impact velocity, indicating uniform deformation.

(c) Larger particles, with higher kinetic energy, require lower velocities to initiate oxide separation and material jetting, leading to the formation of metallurgical bonds at lower velocities.

(d) Soft-to-soft impacts require lower velocities to induce oxide film cracks, resulting in larger interface areas and more oxide-free contact zones. This reduces the critical velocity for soft particles to impact a soft substrate.

(e) The volume of residual oxide for different particle sizes follows a power law equation, with fitting exponents corresponding to particle size scaling parameters in the literature. This suggests that the size effect on the critical velocity is due to the oxide-cleaning ability of particles with varying diameters.

This study highlights the importance of considering oxides in numerical simulations of CS processes to accurately represent deformation and deposition behaviors. It provides valuable insights into the fracture and removal of oxides and the subsequent metallic bond formation during CS processes, offering beneficial new knowledge for the rational design and optimization of CS processes.

**Acknowledgment**

The authors acknowledge the financial support from the Natural Sciences and Engineering Research Council of Canada (Grant #: NSERC RGPIN-2023-03628 and NETPG 493953-16) and the McGill Engineering Doctoral Award (MEDA). The authors would also like to acknowledge the Digital Research Alliance of Canada for providing computing resources.